# Effect of sputter deposited ZnO thin layers on magnetoimpedance response and field sensitivity of Co-based micro-wires

A. Dadsetan [1], M. Almasi Kashi [1,2,*], S.M. Mohseni [3], M. R. Hajiali [3]

[1] Institute of Nanoscience and Nanotechnology 87317, University of Kashan, Kashan, Iran
[2] Department of physics, University of Kashan, Kashan, Iran
[3] Faculty of physics, Shahid Beheshti University, Tehran 19839, Iran

**Abstract**

Semiconductive ZnO has shown to be a fascinating element for application in high performance sensors. ZnO with various thicknesses (87-500 nm) was deposited on the surface of Co-based amorphous micro-wire ($Co_{68.15}Fe_{4.35}Si_{12.5}B_{15}$) using magnetron sputtering technique and the magnetoimpedance (MI) response was evaluated. The MI% increases monotonously up to the 240 nm thickness of ZnO layer. By further increase in the ZnO layer thickness the MI% decreases. MI response and field sensitivity increased from 227% and 1.7 %/Oe for bare micro-wire up to 406% and 4.99 %/Oe for ZnO deposited micro-wire with a thickness of 240 nm. The x-ray diffraction (XRD) and scanning electron microscopy (SEM) were used to characterize the structural grown ZnO layers. Vibrating sample magnetometry (VSM) was used to reveal the role of the ZnO thin layer in the MI response. The same trend of increase and decrease in the MI% by thickness was observed for the transverse permeability of the samples determined by VSM. The obtained results address a simple way to achieve high MI response and sensitivity with selective surface sensing.

[*]Corresponding author's email address: **almac@kashanu.ac.ir**

# 1. Introduction

Magnetoimpedance (MI) effect consists of a significant change in the impedance (both real and imaginary parts) of a soft magnetic conductor under the influence of an external magnetic field [1,2]. This effect is very important due to its capabilities to be used for technological applications such as magnetic field sensors [3]. The MI effect has been studied in different forms of soft magnetic materials such as ribbons, films, micro-wires, and their nanocrystalline counterparts [4,5]. MI is correlated with the skin depth ($\delta$), $\delta = (\rho/\pi\mu_t f)^{1/2}$ of the high frequency $f$ current and the transverse magnetic permeability $\mu_t$ of metallic ferromagnet with electric resistivity $\rho$. $\mu_t$ changes by applying external field and this results in a new current skin depth and so varying the MI response [2].

Magnetic properties of soft magnetic materials and MI effect will be influenced by coating of a magnetic or conducting or insulating layer on bare samples [6–12]. A controlled engineering of surface of a soft ferromagnetic ribbon has proved its useful controllability in enhancing the MI effect and so its practical application in magnetic sensors. Toward this, we can refer to some examples of dielectric materials, which are Zinc oxides [8], Copper oxide [9] diamagnetic organic thin film [10], $CoFe_2O_4$ [11] and very recently graphene oxide [12]. These papers have discussed the influence of the surface structure on the magnetic field and surface domain structure and so tuning the MI response. Also the reports of the conductive magnetic layers coated on the surface of MI sensors [13] have discussed on the magnetoimpedance exchange coupling. Very recently we have proposed that impedance spectroscopy can be used for the detection of spin orbit torque resulting from the spin hall effect in ribbon/Pt [14] and ribbon/IrMn heterostructures [15]. Therefore, surface sensitivity of the MI effect especially when the skin effect is sensitive against any small variations is significant. Besides, annealing and applying tensile stress to ribbons towards improvement of sensitivity is another interesting subject discussed by many authors [16–18].

The studied heterostructure in this article is made of a ferromagnetic Co-based micro-wires and a thin layer of semiconductor ZnO. Among all oxide semiconductor materials, ZnO has been considered as an important element in technological devices due to its wide applications in photodetectors [19], solar cells [20], biosensors or gas sensors [21,22], and spintronic devices [23]. Furthermore, room temperature ferromagnetism has been found in undoped ZnO [24–26] in the forms of thin films, which is attributed to oxygen defects, especially singly ionized oxygen vacancies.

There are different deposition techniques used for the preparation of ZnO thin films such as laser deposition [27], sol-gel [28], magnetron sputtering [29], chemical vapor deposition [30] and electrodeposition [31]. Among these, sputtering offers advantages such as low temperature of fabrication while yielding preferred crystalline orientation and uniform properties. Based on only one study, ZnO thin film was deposited on a magnetic ribbon using SILAR method that resulted in 95.37% improvement (8 to 15.63%) in the MI effect [8]. The thickness for the deposition of ZnO layer was within micrometer range which resulted in a low MI effect and further accomplishment in this study is required. In this paper, we systematically study the effect of different thicknesses of sputter deposited ZnO thin films on the MI response and field sensitivity of co-based micro-wires. The highest MI ratio of 406% and the highest field sensitivity of 4.99 %/Oe are obtained at ZnO thickness of 240 nm. Vibrating sample magnetometry (VSM) hysteretic plots and consequently derived permeabilities are successively used to interpret the effect of ZnO thin film layers on MI response. Results are considerable in development of highly sensitive MI sensors.

## 2. Experimental

Amorphous micro-wires with the nominal composition of $Co_{68.15}Fe_{4.35}Si_{12.5}B_{15}$ (42 mm in length, effective length of 40 mm, and 180 μm in diameter) were used the samples were initially cutting into suitable dimension and then were immersed in Dimethyl sulfoxide polishing solution and sonicated for 30 minutes. After that, they were immersed in a beaker containing distilled water for 15 minutes. Finally, the samples were baked in an oven for 20 minutes in a temperature of 60 °C to dry and be ready for the deposition process. ZnO layer was deposited on the cleaned Co-based micro-wire samples using magnetron sputtering technique. The base pressure and argon pressure was $1.5\times10^{-7}$ and $1.5\times10^{-2}$ mbar, respectively. ZnO layer was deposited using a ZnO target (99.999%) with thicknesses ranging from 87 to 500 nm, as shown schematically in Fig.1. X-ray diffraction (XRD) was performed using a STADI STOE diffractometer with CuKα ($\lambda$ = 1.54 Å) for the angle (2$\theta$) range from 10–80°. Transverse magnetization curves of the samples were measured using a VSM (500 MDK). To measure the MI response of samples, an external magnetic field produced by a solenoid applied along the micro-wire axis and the impedance was measured by means of the four-point probe method. An ac current passed through the longitudinal direction of the micro-wire with different frequencies supplied by function generator (GPS-2125), with constant amplitude of current. The impedance was evaluated by measuring the voltage and current across the sample using a digital oscilloscope (GPS-1102B). The MI ratio can be defined as

$$MI\% = \frac{Z(H)-Z(H_{max})}{Z(H_{max})} \times 100 \quad (1)$$

In which $Z$ is the impedance as a function of external field $H$, and $H_{max}$ is the maximum field applied to the samples in the MI measurement. Magnetic field sensitivity is defined as the slope of the MI curve of the sample

$$S = d\left(\frac{\Delta Z}{Z}\right)/dH . \quad (2)$$

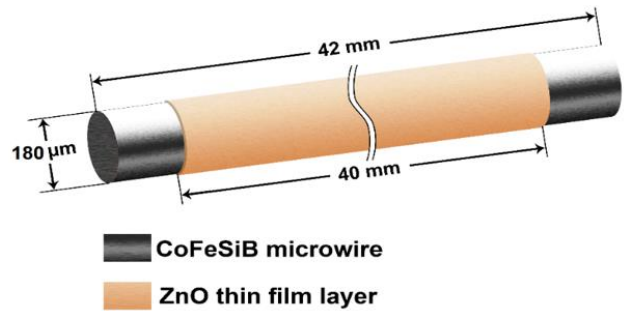

Fig. 1. Schematic diagram of CoFeSiB micro-wire coated with ZnO thin film.

## 3. Results and discussion

Fig. 2 shows the XRD pattern of bare micro-wire and 240 nm ZnO deposited micro-wire. The XRD pattern of bare micro-wire shows an amorphous nature. For 240 nm ZnO deposited micro-wire, the ZnO layer is crystalline in nature and the coexistence of two peaks at $2\theta = 34.39°$ and $36.18°$ shows the formation of (002) and (101) crystalline planes.

To investigate the effect of the deposition of the ZnO layer on the magnetic properties of the micro-wires, the magnetic hysteresis loops of the undeposited and ZnO deposited layer with various thicknesses were recorded at room temperature. Figure 3(a-d) shows the magnetic hysteresis loops of undeposited and ZnO-deposited (87, 215 and 240 nm) micro-wires normalized to the saturation value ($M_S$). The loops are all thin and narrow, and magnetization is saturated at a small applied field, indicating their soft ferromagnetic characteristics. There is just a little difference between deposited and undeposited micro-wires. This is because of the large volume of micro-wires has a dominant contribution in the measured magnetization of the samples. There are indeed changes in coercivity ($H_c$) and slope at zero point (permeability) which were determined for all the samples as shown in the Fig 4(e,f).

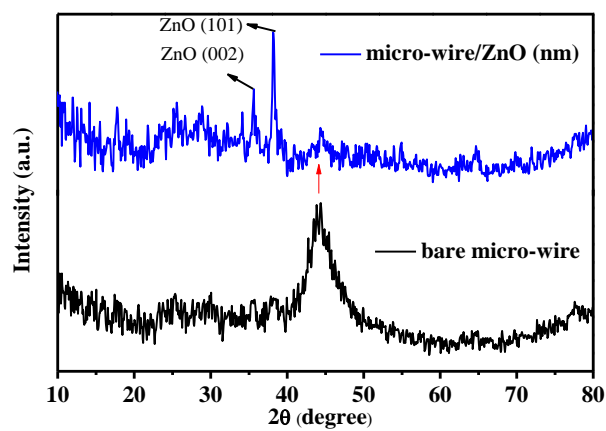

Fig. 2. X-ray diffraction pattern of bare micro-wire and 240 nm ZnO deposited

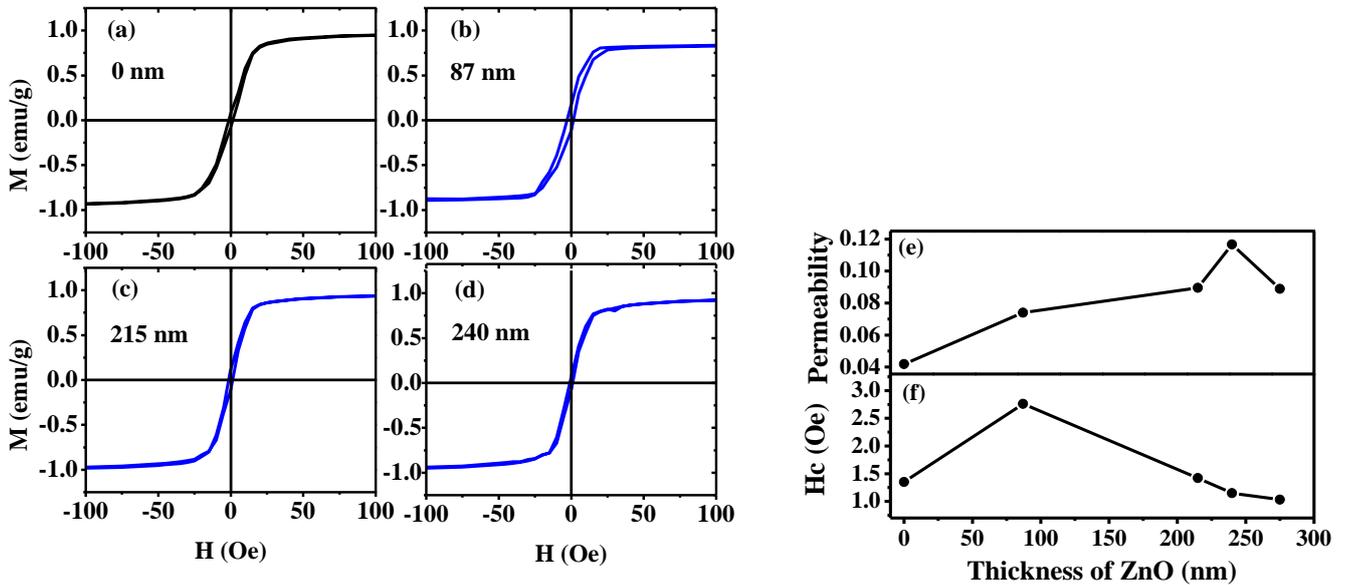

Fig. 3. Normalized magnetization curves for ZnO deposited layer with thickness of (a) 0, (b) 87, (c) 215, (d) 240 nm and effect of thickness of ZnO layer on (e) the slope at zero point (permeability) (f) coercivity ($H_c$).

From the hysteresis loops of Fig. 3(a-d) it is observed that the coercivity ($H_c$) decreases by increasing the thickness of ZnO layer. Nevertheless, the decisive parameter determining the MI% of the samples is the $\mu_t$. Therefore, the $\mu_t$ of the samples were derived (Fig. 3e). It is observed that the $\mu_t$ increases by increasing the thickness of the ZnO layer up to 240 nm. By further increase in the thickness the $\mu_t$ decreases too. This result can be used in interpretation of the MI results.

Scanning electron microscopy (SEM) images of the bare micro-wire and 240 nm ZnO-deposited micro-wires are presented in figures 4(a,b), respectively. As can be seen, ZnO is formed in a granular structure atop the micro-wire surface. The surface if fully covered by the layer, a common characteristic of the layers fabricated by the sputtering method. As known, surface coverage of MI sensors without voids is vital for an effectively enhanced MI% with surface selectivity.

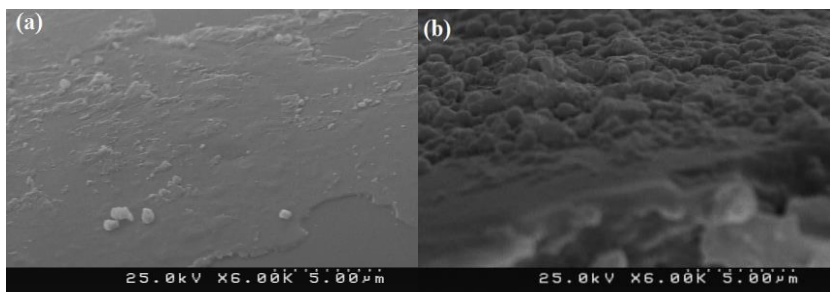

Fig. 4. SEM images of (a) the bare micro-wire and (b) 240 nm ZnO thin layer deposited on micro-wire.

In order to understand the impact of ZnO deposition on the MI response of the micro-wire, magnetic field and frequency dependences of MI ratio of all the samples were measured up to 300 Oe, magnetic field strength and at different frequencies (1-10 MHz). Figure 5 (a) shows the MI ratio of the bare and deposited micro-wire with different thickness of ZnO at $f$ = 2 MHz. As can be seen in Fig. 5 (a), MI response for ZnO deposited samples changes significantly. Increasing the thickness of the deposited layers from 87 to 240 nm, MI response increases considerably and then decreases by further increase in thickness. In figure Fig. 5 (c,d), the maximum of MI ratio and field sensitivity for different thicknesses of ZnO layers at $f$ = 2 MHz can be seen. Maximum MI ratio of ~406.64% is obtained for the sample with the 240 nm ZnO layer. To better illustrate the effect of ZnO deposition, Maximum of the MI response for the bare and deposited micro-wire with 240 nm at various frequencies (1–10 MHz) are presented in Fig. 4(b). It is noted that with increasing frequency, the maximum MI ratio initially increases, reaches a maximum at a particular frequency (2 MHz), and then decreases for higher frequencies. Relative contributions of domain wall motion and magnetization rotation to the $\mu_t$ should be considered in interpreting this trend [3]. The reduction of MI ratio at high frequencies is due to presence of eddy currents that causes damping of domain wall displacements and only rotation of magnetic moments takes place. In turn the $\mu_t$ diminishes, and the MI ratio decreases. The increase of MI response in the magnetic layer deposited on ribbon was already explained according to modifications of the ribbon surface and the closure of magnetic flux paths of deposited ribbons with magnetic materials having low thicknesses and decreasing MI for higher thicknesses [13]. We see the same trend and the MI ratio for ZnO deposited changes accordingly. These effects as well as the physical aspects of thickness and layer dependence of MI response according to roughness, and phase of ferromagnetic turns into diamagnetic by increasing the thickness of deposited layers is discussed. As ZnO is a well-known oxide semiconductor, we can conclude that there is less current passing through it and therefore no change in the skin depth of the micro-wire does occur. In the case of our samples, the ZnO oxide semiconductor layer diminishes the magnetic field passing through the surface. Therefore, the magnetic field penetrates better inside the magnetic micro-wire relative to the bare micro-wire. Also, the measured improvement in the MI response of the deposited micro-wire by ZnO can be considered as a result of the stress on the domain structure of the micro-wire and changing its magnetization dynamics [9].

According to the Fig. 3, another reason for enhancement of MI response is increase of magnetic permeability of the ZnO deposited sample which is proportional to the magnetization slope. Since the MI effect strongly depend on $\mu_t$, by increasing $\mu_t$ the MI effect increases.

As mentioned above, there are many reports on the increase of MI% by coating ferromagnetic layer on the surface of MI sensors [6-12]. Therefore, for the explanation of the evolution of the MI response we take look at the magnetic characteristics of ZnO reported by others [29]. As can be seen in fig. 4(a) the MI response significantly decreases with an increase in thicknesses of ZnO layer more than 240 nm. Kapilashrami et al. reported a systematic study of the film thickness dependence (0.1-1 $\mu m$) of room-temperature ferromagnetism in pure magnetron-sputtered ZnO thin films [29]. They observed that the ZnO films exhibit a sequential transition from ferromagnetism to paramagnetism and diamagnetism as a function of film thickness. We therefore speculate that by increasing those thin layer thicknesses above the 240 nm, ZnO thin layer evolved from ferromagnetism to paramagnetism and thereby MI effect decreases. It is worth mentioning that the thicknesses of the layers are well above the thresholds which their magnetic properties could be affected by the substrate. Therefore, our speculation for the similar magnetic evolution is reasonable.

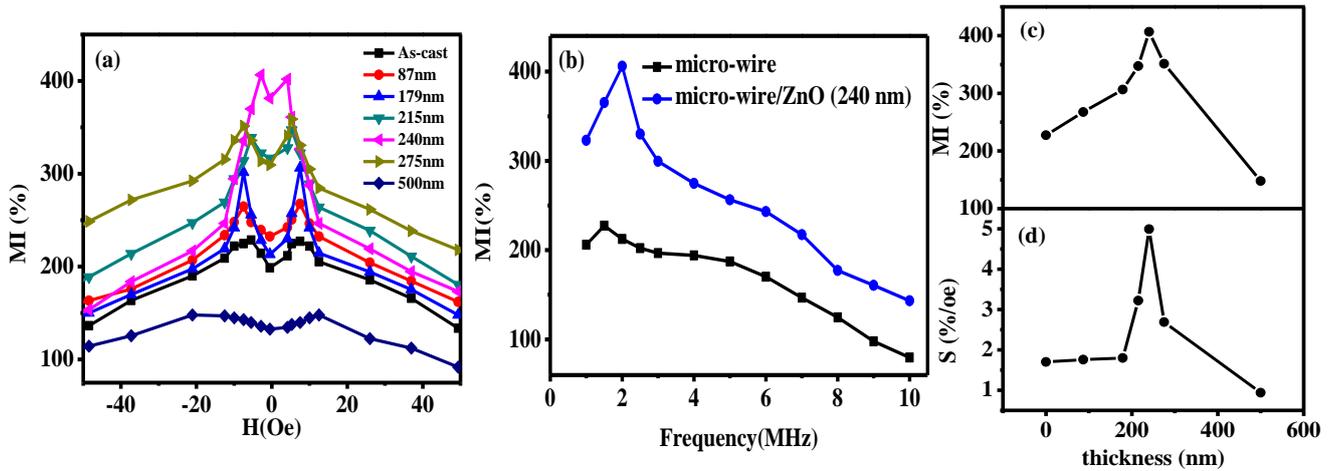

Fig. 5. (a) MI response for bare micro-wire and ZnO deposited with various thicknesses at $f$= 2 MHz. (b) Maximum MI response for bare micro-wire and ZnO (240 nm) deposited versus frequency. (c,d) Maximum MI response and field sensitivity versus thickness for ZnO deposited samples at $f$ = 2 MHz.

## 4. Conclusion

ZnO thin layer deposited by magnetron sputtering technique (87-500 nm), improved MI response and field sensitivity of Co-based micro-wire up to 406% and 4.99%/Oe respectively. The evolution of MI% was well understood by the magnetic properties of the ZnO layer at different thicknesses. The transverse permeability determined by VSM results, was used to reveal the role of the ZnO thin layer in the MI response and field sensitivity. XRD and SEM showed the granular structure of ZnO layer with (101) and (200) crystalline aligned directions. The result of this study show that, thin dielectric layers is one of the easiest and at the same time the most efficient methods for MI response optimization; and sensing elements made by this method can be a suitable candidate for development of high-performance MI sensors.